\documentclass[12pt,3p]{elsarticle}
\usepackage{amssymb}
\usepackage{amsmath}
\usepackage{amsfonts,amsthm,bm} 
\usepackage{graphicx}
\usepackage{caption}
\usepackage{hyperref}
\usepackage{color, colortbl}
\definecolor{silver}{rgb}{0.75,0.75,0.75}
\usepackage{lscape}
\usepackage{rotating}
\usepackage{diagbox}
\usepackage{subcaption}
\usepackage{csquotes}
\usepackage{floatpag}
\usepackage{setspace}
\PassOptionsToPackage{hyphens}{url}\usepackage{hyperref}
\makeatletter
\def\ps@pprintTitle{%
\let\@oddhead\@empty
\let\@evenhead\@empty
\def\@oddfoot{\centerline{\thepage}}%
\let\@evenfoot\@oddfoot}
\makeatother

\usepackage{etoolbox}
\patchcmd{\MaketitleBox}{\footnotesize\itshape\elsaddress\par\vskip36pt}{\footnotesize\itshape\elsaddress\par\parbox[b][36pt]{\linewidth}{\vfill\hfill\textnormal{\today}\hfill\null\vfill}}{}{}%
\patchcmd{\pprintMaketitle}{\footnotesize\itshape\elsaddress\par\vskip36pt}{\footnotesize\itshape\elsaddress\par\parbox[b][36pt]{\linewidth}{\vfill\hfill\textnormal{\today}\hfill\null\vfill}}{}{}%

\newcommand{\CE}[1]{{{}}}

\begin{document}
\begin{frontmatter}

\renewcommand*{\thefootnote}{\fnsymbol{footnote}}

% % % anonymous version
%\title{Vulnerable Funds?}
%\date{---}

 % % actual version
 \title{ Stress testing and systemic risk measures using elliptical conditional multivariate probabilities }
\author{Tomaso Aste}
\address{Department of Computer Science, University College London, \\ Gower Street,  WC1E 6EA, London, United Kingdom;\\
UCL Centre for Blockchain Technologies, UCL, London, UK;\\
Systemic Risk Centre, London School of Economics and Political Sciences, London, UK. }

\begin{abstract}
Systemic risk, in a complex system with several interrelated variables, such as a financial market, is quantifiable from the multivariate probability distribution describing the reciprocal influence between the system's variables.
The effect of stress on the system is reflected by the change in such a multivariate probability distribution, conditioned to some of the variables  being at a given stress' amplitude. 
Therefore, the knowledge of the conditional probability distribution function can provide a full quantification of risk and stress propagation in the system. 
However, multivariate probabilities are hard to estimate from observations. 
In this paper, I investigate the vast family of multivariate elliptical distributions, discussing  their estimation from data and proposing novel measures for stress impact and systemic risk in systems with many interrelated variables. 
Specific examples are described for the multivariate Student-t and the multivariate normal distributions applied to financial stress testing. 
An example of the US equity market illustrates the practical potentials of this approach.
\end{abstract}
\begin{keyword}
Stress testing, Systemic risk, Elliptical conditional probability  \\
% \textit{JEL classification}: % G10; G11; G23
\end{keyword}

\end{frontmatter}

% \newpage

\section{Introduction}

\renewcommand*{\thefootnote}{\arabic{footnote}}
\setcounter{footnote}{0}

\onehalfspacing
In financial systems, stress testing consists in quantify the ability of the system to cope with a crisis and identify weaknesses. 
In particular, it is important to identify weaknesses that can affect a large part of the system and they are therefore sources of systemic risk.  Such tests are important to assess the robustness of the system and they are required by market regulators. 
Besides finance, stress testing and systemic risk assessment are crucial tasks in many other domains such as ecological or biological systems. 
These are complex systems where many interdependent variables are defining the system state and its dynamics.
The most general approach for stress testing  consists in estimating the effects of setting some variables (the `stressing' variables $\mathbf{X}$) at extreme values onto the statistical properties of another set of variables (the `stressed' variables $\mathbf Y$). 
This requires the computation of the conditional probability $P(\mathbf Y|\mathbf{X})$. 
The computation of such a conditional probability can be challenging in multivariate systems comprising a large number of variables, such as financial markets. 
However, for a vast set of probability distributions that belong to the elliptical family, the challenge is mainly reduced to estimate, with accuracy, the covariance matrix \cite{fang2018symmetric}.  
In this paper, I report how, for this probability distribution family, one can introduce practical multivariate systemic risk measures, useful for stress testing and quantification of risk \cite{feinstein2017measures}.

There is a vast literature on systemic risk in financial systems (see for instance in \cite{acharya2017measuring} for an overview); where, through different approaches, researchers study the propagation and amplification of losses  in markets caused by externalities that can trigger shortfalls when the system is undercapitalized. Stress on an industry sector can cause fire-sales and trigger externalities on other institutions and different industry sectors. Spillover effects can propagate and they can be amplified through the system causing distortions and  systemic effects that eventually can involve the whole system compromising its stability. They are therefore called systemic. 
To ensure the stability of the system, it is important to be able to quantify the propagation of stress through the system from institution to institution across sectors.  
Several methodologies have been proposed to quantify systemic risk with approaches ranging from general economic perspectives \cite{allen2000financial} to network theoretical approaches \cite{birch2014systemic,caccioli2018network,tungsong2017relation}, specific econometric tools \cite{biagini2019unified,billio2012econometric}, machine learning techniques \cite{kou2019machine} and game-theoretic reasoning \cite{tarashev2009systemic}.
Overall, all these approaches are aiming at capturing the propagation of stress and consequently losses through the system \cite{alexander2008developing,bensalah2000steps,de2008stress,longin2000value,pritsker2019overview,rebonato2010coherent} and assess the criticality of the system identifying specific fragilities. 
This task requires the modeling of the complex set of variables and mechanisms characterizing the highly interconnected system of financial institutions, industries and banks at the basis of our economy.
This paper adds on to the existing literature by introducing measures for stress propagation and systemic risk that are truly multidimensional quantifying directly effects between sets of variables and not just couples of them. 
These measures are general for the vast family of elliptical distributions and can be extended further, for instance including the multivariate generalized hyperbolic class \cite{Prause99}. 

Mathematically, this is a system comprised of many dependent variables that can be statistically described by a joint probability distribution. 
The risk in the system is measured by the probability of negative fluctuations (losses). 
The systemic risk is associated with the interdependency between the variables in the system and therefore with the probability of occurrence of several losses across different institutions.    
Propagation of stress is instead associated with the conditional probability of a loss given that some other variables are stressed at a certain amount of loss.  
This is the approach adopted in  \cite{adrian2011covar,adrian2016covarPuB} where a conditional definition of value at risk, named CoVaR, was introduced.
Such a conditional value at risk is the value at risk (VaR) of the `stressed' variable, $Y$, computed conditioned on the  `stressing' variable $X$ being at its VaR value for a certain level of risk. 
This is an instance of the use of conditional probability to compute a measure of risk when two variables are involved.  
Two major limits of the CoVar approach are that it applies to only two variables and it measures risk in terms of VaR.
Conversely,  in these system collective phenomena play a very important role and  VaR is only one of many possible risk quantification measures.  

Stress applied to a subset of  variables does not only change the quantile (VaR) of the other variables, rather it changes both location and shape of the multivariate distribution of the stressed variables.
In this paper I argue that the most important contribution to risk in a system under stress is caused  by the shift of the centroid of the conditional probability distribution with respect to the centroid of the unconditioned probability distribution. For markets, such a shift can be quantified in money value and interpreted as  propagation of stress.

The value of such a shift is the same for all elliptical distribution and it is therefore rather general. 
For this family of probability distributions, the other important contributions to systemic risk are the change in the shape of the distribution and the rotation of the principal axis of the equi-probability ellipsoid. 
In this paper I present three measures of  systemic risk for multivariate systems under stress which are associated to these three contributions. 
These measures are general for the whole family of elliptical distributions. 
I also introduce some other measures in the appendix and discuss their relations with the previous.

This paper is organized as follows: in Section \ref{s.1}, to set the tone and introduce some important concepts, 
I briefly report about the CoVaR reasoning, adopting however a slightly different perspective with respect to the original paper. Expressions for the conditional multivariate probabilities for the whole elliptical distribution family are obtained in Section \ref{s.2} with a detailed discussions for the multivariate Normal and multivariate Student-t distributions. In Section \ref{s.2a} the effect of conditioning on the properties of elliptical distributions is discussed and systemic risk measures and stress testing tools from these conditional probabilities are introduced. 
In Section \ref{s.4}, I discuss an example concerning the study of systemic impacts between industry sectors under stress from the analysis of equity returns in the US market.  
Conclusions and perspectives are given in Section \ref{s.5}.
Three sections in the \ref{A.VariabilityFactors}, introduce other measures of stress testing and systemic risk and discuss their relation with the ones introduced in section  \ref{s.2a} .
\ref{A.b} reports on the application of these measures on the US returns dataset comparing the results with the ones in Section \ref{s.4}.

\section{Univariate measure of risk in terms of Value at Risk and Conditional Value at Risk}
\label{s.1}

\subsection{ Value at Risk ({VaR})}
For continuous variables $\mathrm{VaR}(q)$  is  the $q$-quantile of the loss distribution, meaning that with probability $q$, losses will not exceed the value $\mathrm{VaR}(q)$. 
Specifically, for a random variable $X$ representing the losses, $VaR_{ X}(q)$ is defined by
\begin{equation}
P(X \le \mathrm{VaR}_{ X}(q)) = q.
\label{VaR}
\end{equation} 
When the random variable $X$ belongs to the location scale family, then it has the property that the distribution function of the scaled, shifted variable $a + b X$ also belongs to the same family.
Therefore, I can arbitrary scale the variable while preserving the distribution and in particular I can always write the distribution with respect to the standard form:
\begin{equation}
P(X \le \mathrm{VaR}_{ X}(q)) = P\left[\frac{X- {\mu}_{ X}}{\sigma_{ X}} \le  \frac{\mathrm{VaR}_{ X}(q)- {\mu}_{ X}}{\sigma_{ X}} \right] = \Phi_{X}\left[\frac{\mathrm{VaR}_{ X}(q)-\mu_{ X}}{\sigma_{ X}} \right] = q ,
\end{equation}
where $\mu_{ X}$ is the location parameter, $\sigma_{ X}$ is the scale parameter and  $ \Phi_X( \cdot)$ is the standard form of the cumulative distribution function associated with the statistics of the standardized variable. 
If the inverse of such cumulative distribution exists, then
\begin{equation}
\frac{\mathrm{VaR}_{ X}(q)-\mu_{ X}}{\sigma_{ X}} =  \Phi_X^{-1}(q).
\end{equation}
and therefore I have a general expression for the $\mathrm{VaR}(q)$:
\begin{equation}
\mathrm{VaR}_{ X}(q) =  \mu_{ X} + \Phi_X^{-1}(q) \sigma_{ X} \;\;.
\label{VaR1}
\end{equation}

\subsection{Conditional Value at Risk ({CoVaR})}

In  \cite{adrian2011covar,adrian2016covarPuB} a conditional version of the $\mathrm{VaR}_{ X}(q)$ was proposed as a measure of impact of a variable on the risk of another. 
In analogy with the definition of VaR from Eq.\ref{VaR} the conditional VaR is 
\begin{equation}
P(Y \le \mathrm{VaR}_{ Y| X}(q) | X) = q.
\label{cVaR}
\end{equation} 
Which is identical to the definition of $\mathrm{VaR}_{ X}$ in Eq.\ref{VaR} except that the probability is now a conditional probability. {\color{black}
In analogy with Eq.\ref{VaR1} the conditional VaR is 
\begin{equation}
\mathrm{VaR}_{ Y| X}(q) =  \mu_{ Y| X} + \Phi_{Y|X}^{-1}(q) \sigma_{ Y| X} \;\;.
\label{VaRcondit}
\end{equation}
where $ \Phi_{Y|X}(\cdot)$ is the  standard form of the cumulative distribution function associated with the statistics of the standardized variable, it can be different from $ \Phi_{X}(\cdot)$ due to conditioning that can change the parameters of the distribution (for instance, this is the case for the Student-t, as we shall see).
In  \cite{adrian2011covar,adrian2016covarPuB}  the value of $X$ is stressed at its VaR value $X = \mathrm{VaR}_{X}(q)$.

For the  location scale family the conditional probability has the same form as the unconditional probability but with shifted location parameter and modified scale parameter.
For the whole location scale family  the location parameter under conditioning becomes:
\begin{equation}
\mu_{ Y| x} = \mu_{Y} + \rho_{X,Y}\frac{\sigma_{ Y}}{\sigma_{ X}}(x-\mu_{ X})
\label{MY|Y}
\end{equation}
where $\rho_{X,Y}$ is the Pearson correlation coefficient between variable $X$ and variable $Y$. 
Note that this shifted location parameter $\mu_{ Y| x}$ is the least squares linear regression of $Y$ given $X=x$.

The scale parameter change depends on the distribution. Let me here report explicitly for the case of the Normal and Student-t distributions.

\subsection{Normal distribution}
For the normal distribution the scale parameter becomes:
\begin{equation}
\sigma_{ Y| X}^2 = (1-\rho_{X,Y}^2) \sigma_{ Y}^2,
\label{SY|Y}
\end{equation}
Eqs.\ref{MY|Y} and \ref{SY|Y} can be quite straightforwardly retrieved with elementary algebra by completing the square in the expression of the joint distribution considering $X=x$ as a constant and shifting $Y$.
Essentially, for the normal case the probability distribution shifts its location by the factor in Eq.\ref{MY|Y} and changes the scale by a factor $\sqrt{1-\rho_{X,Y}^2}$.
This also implies that any increase in the value at risk of the conditioned variable is only driven by the linear shift because the variance of $Y$ is reduced by  the conditioning. 
Some traditional risk measures and CoVaR applications for systemic risk can be found in \cite{kurosaki2013systematic}.

\subsection{ Student-t}
In the case of Student-t distribution we have the same shift in the location parameter as in Eq.\ref{MY|Y}. However, the degrees of freedom increase to $\nu+1$  and the scale parameter changes to:
\begin{equation}
\sigma_{ Y| x}^2 = \frac{\nu+d^2_x}{\nu+1}(1-\rho_{X,Y}^2) \sigma_{ Y}^2.
\label{SY|Yt}
\end{equation}
with $d^2_x = (x-\mu_x)^2/\sigma_X^2$.
Contrary to the normal case, in the Student-t case the scale factor depends on the values of $X=x$. 
Therefore, the scale of the distribution of $Y|X=x$ around the shifted centre $\mu_{ Y|x}$ can be smaller or larger than the scale of the unconditional $Y$. 
For instance, when the system is not stressed and $x=\mu_X$, then the conditional scale is smaller than the unconditional one. However, when the stress is deviating from the mean more than one standard deviation, then the conditional scale becomes larger than the unconditional.
This has relevant consequences for any risk measure which can increase substantially beyond the linear shift of the conditional mean.

In this non-linear case, due to the dependence of $\sigma_{ Y| x}$ from the values of $X=x$, the residual $Y-\mu_{ Y|X}$ is not independent from $X$.
However, its scaled counterpart  $ \frac{Y-\mu_{ Y|X}}{\nu+d^2_X}$ is instead independent from $X$.\footnote{Note that we consistently used lower case $x$ to indicate the value of the random variable $X$. Consistently, $d^2_x$ is a number, whereas $d^2_X$ is instead a random variable.}

}

\section{Conditional probability for the elliptical family} \label{s.2}
Let us now extend the reasoning illustrated in the previous Section to a multivariate case with $\mathbf{X} \in \mathbb R^{p_{\mathbf X} \times 1}$ and $\mathbf{Y} \in \mathbb R^{p_{\mathbf Y}\times 1}$. 
I consider the case when the multivariate probability distribution of the random variables $\mathbf{Z} = (\mathbf{X}^\top,\mathbf{Y}^\top)^\top \in \mathbb R^{p_{\mathbf Z} \times 1}$  (with $p_{\mathbf Z} = p_{\mathbf X}+p_{\mathbf Y}$) belongs to the elliptical family, which implies that the multivariate probability density function can be written as 
\begin{equation}
f_{\mathbf Z}(\mathbf{Z}) 
= k  |\mathbf{\Omega}|^{-1/2}  g(d^2_\mathbf{Z})
\label{fZ}
\end{equation}
where $g(\cdot)$ is a scalar function which is the standardized form of the distribution, its expression is independent from the location and scale parameters but dependent on dimension $p_{\mathbf Z}$ and eventually on other parameters (such as the degrees of freedom $\nu$ in the Student-t case), $k$ is a constant that also can depend on these parameters. 
The multivariate normal has $g(d^2_\mathbf{Z}) = \exp(-d^2_\mathbf{Z}/2)$ and the multivariate Student-t has $g(d^2_\mathbf{Z})=(1+d^2_\mathbf{Z}/\nu)^{-(\nu+p_{\mathbf Z})/2}$ with $\nu$ the degrees of freedom. 
The matrix $ \mathbf{\Omega} \in \mathbb R^{p_{\mathbf Z} \times p_{\mathbf Z}}$ is a positive definite matrix, which is equal or proportional to the shape matrix and  equal or proportional to the covariance matrix when it is defined. 
The quantity $d^2_\mathbf{Z}$ is the scalar:
\begin{equation}
d^2_\mathbf{Z} = (\mathbf{Z}- \boldsymbol{\mu}_{\mathbf Z})^\top  \mathbf{\Omega}^{-1} (\mathbf{Z}- \boldsymbol{\mu}_{\mathbf Z}).
\label{fd2}
\end{equation}
which is the square Mahalanobis distance \cite{chandra1936generalised} when $ \mathbf{\Omega}$ is the covariance matrix, as in the case of the multivariate normal distributions, and it is its generalization in all other cases. 
Here, $ \boldsymbol{\mu}_{\mathbf Z} \in \mathbb R^{p_{\mathbf Z} \times 1}$ is the position of the centroid of the equi-probability elliptical surface and it is given by the vector of means or location parameters of the marginals.
The symbol $|\cdot|$ indicates the determinant.

%%%%%%%%%%%%%%%%%%%%%%%%%%%%%%%%%%%%%%%%%%%
{\color{black}
When the probability density function is defined, the multivariate conditional probability of $\mathbf Y \in \mathbb R^{p_{\mathbf Y}\times 1}$ given $\mathbf X \in \mathbb R^{p_{\mathbf Y} \times 1}$ is defined through the Bayes formula 
\begin{equation}
f_{\mathbf Y | \mathbf X}(\mathbf Y) = \frac{f_{\mathbf X \mathbf Y}( \mathbf X, \mathbf Y )}{f_{\mathbf X}( \mathbf X )} .
\end{equation}
When conditioning, one variable is fixed to some values, $ \mathbf X =  \mathbf x$, therefore $f_{\mathbf X}( \mathbf X =  \mathbf x )$ in the previous formula is a constant and consequently 
\begin{equation}
f_{\mathbf Y | \mathbf x}(\mathbf Y) \! \propto \! f_{\mathbf X \mathbf Y}( \mathbf X= \mathbf x, \mathbf Y)  \! \propto \! g(d^2_{\mathbf Y | \mathbf x} + d^2_{\mathbf x} ) .
\end{equation}
with
\begin{equation}
d^2_{\mathbf Y |\mathbf x} = 
({\mathbf Y}-{\boldsymbol\mu_{\mathbf Y| \mathbf x} })^\top
\boldsymbol \Omega_{\mathbf Y\mathbf Y| \mathbf X}^{-1}
({\mathbf Y}-{\boldsymbol\mu_{\mathbf Y|\mathbf x} })
\end{equation}
and
\begin{equation}\label{d2x}
d^2_{\mathbf x} = 
({\mathbf x}-{\boldsymbol\mu_{ \mathbf X} })^\top
\boldsymbol\Omega^{-1}_{\mathbf X\mathbf X}
({\mathbf x}-{\boldsymbol\mu_{\mathbf X} }),
\end{equation}
with $ \mathbf{\Omega}_{\mathbf X \mathbf X} $ assumed to be invertible. 
Note that for any given $\mathbf X= \mathbf x$, the term $d^2_{\mathbf x}$ is a constant.
The other terms are: 
\begin{equation}\label{f.CondMU}
\boldsymbol\mu_{\mathbf Y| \mathbf x} =  \boldsymbol{\mu}_{\mathbf Y} +  \mathbf{\Omega}_{\mathbf{YX}} \mathbf{\Omega}^{-1}_{\mathbf{XX}} (\mathbf{x}- \boldsymbol{\mu}_{\mathbf X}),
\end{equation}
 the vector of conditional centroids; and 
\begin{equation}\label{f.CondOM}
\boldsymbol \Omega_{\mathbf Y\mathbf Y| \mathbf X} 
= \boldsymbol{\Omega}_{\mathbf Y \mathbf Y} -   \boldsymbol{\Omega}_{\mathbf Y\mathbf X} \boldsymbol{\Omega}^{-1}_{\mathbf X\mathbf X}  \boldsymbol{\Omega}_{\mathbf X \mathbf Y},
\end{equation}
which is the inverse of the Shur complement $ (\boldsymbol{\Omega}^{-1})_{\mathbf Y \mathbf Y} $. 
The terms $ \mathbf{\Omega}_{\mathbf X \mathbf X} $, $ \mathbf{\Omega}_{\mathbf X \mathbf Y}$, $ \mathbf{\Omega}_{\mathbf Y \mathbf Y}$ and $ \mathbf{\Omega}_{\mathbf Y \mathbf X}$ are the block elements of the  joint shape matrix $ \mathbf{\Omega}$
\begin{equation}
 \mathbf{\Omega} = 
\left(\begin{array}{cc}
 \mathbf{\Omega}_{\mathbf X \mathbf X}& \mathbf{\Omega}_{\mathbf X \mathbf Y}\\
 \mathbf{\Omega}_{\mathbf Y \mathbf X}&  \mathbf{\Omega}_{\mathbf Y \mathbf Y}
\end{array}\right).
\end{equation}

Summarizing, the conditional probability density function (when is defined) for any probability of the elliptical family with $f_{\mathbf Z} \! = \! k  |\mathbf{\Omega}|^{-1/2}  g( d^2_{\mathbf Z } )$ is in the form: 
{\color{black}
\begin{equation}
f_{\mathbf Y | \mathbf x}(\mathbf Y) \! = \! \bar k |\boldsymbol \Omega_{\mathbf Y\mathbf Y| \mathbf X}  |^{-1/2}  g( d^2_{\mathbf Y | \mathbf x} + d^2_{\mathbf x}  ).
\label{e.fYx}
\end{equation}
}
with $\bar k$ a normalization constant that does not depend on $\mathbf Y$.
The functional form of the conditional probability is therefore very similar to the form of the joint beside the additive constant $d^2_{\mathbf x} $ in the argument. 
It is clear that $g( d^2_{\mathbf Y | \mathbf x} + d^2_{\mathbf x}  )$ can always be written as $\tilde g( d^2_{\mathbf Y | \mathbf x} )$ and therefore this conditional probability density function is still a member of the elliptical family.
}
{\color{black}
For all these conditional distributions, the expected values are
\begin{equation}
\mathbb E [ \mathbf Y | \mathbf X=\mathbf x ] = \boldsymbol\mu_{\mathbf Y| \mathbf x} ,
\end{equation}
and the shape matrix is equal or proportional to $\boldsymbol \Omega_{\mathbf Y\mathbf Y| \mathbf X} =\boldsymbol{\Omega}_{\mathbf Y \mathbf Y} -   \boldsymbol{\Omega}_{\mathbf Y\mathbf X} \boldsymbol{\Omega}^{-1}_{\mathbf X\mathbf X}  \boldsymbol{\Omega}_{\mathbf X \mathbf Y}$.
The coefficient of proportionality depends on the distribution and it can be affected by the constraining. 
For what concerns risk, the relevant factor  is the relation between $\boldsymbol \Omega_{\mathbf Y\mathbf Y| \mathbf X}$ and the covariance (when defined). 
Indeed, it is the covariance which quantifies the fluctuations of the variables around the mean.
Hereafter, I discuss in further details the cases for the multivariate normal and the multivariate Student-t distributions.
}

\subsubsection{Conditional distribution for the multivariate normal }
The multivariate normal case is particularly simple because the function 
\begin{equation}
g(d^2_{\mathbf Z })=  \exp(-d^2_{\mathbf Z }/2)
\end{equation}
is an exponential  and therefore the additive constant term, $d^2_{\mathbf x}$ introduces an irrelevant multiplicative constant in front, yielding to 
\begin{equation}
f_{\mathbf Y | \mathbf x}(\mathbf Y) \! = \! \tilde k  |\boldsymbol \Omega_{\mathbf Y\mathbf Y| \mathbf X}  |^{-1/2}  \exp(-d^2_{\mathbf Y | \mathbf x}/2)
\end{equation}
which is a multivariate normal with expectation values $ \boldsymbol\mu_{\mathbf Y| \mathbf x}$ and conditional covariance 
\begin{equation}\label{e.Snorm}
\boldsymbol \Sigma^{nor.}_{\mathbf Y\mathbf Y| \mathbf x} 
= \boldsymbol \Omega_{\mathbf Y\mathbf Y| \mathbf X} .
\end{equation}
Notably, the conditional covariance is different from the unconditional, it is shaped by the correlations between the conditioning and conditioned variables but it does not depend on the value of the conditioning variable $\mathbf X = \mathbf x$. 

\subsubsection{Conditional distribution for the multivariate Student-t }
For the multivariate Student-t  case we have 
\begin{equation}\label{g}
g(d^2_{\mathbf Z })=(1+d^2_{\mathbf Z }/\nu)^{-\frac{\nu+p_{\mathbf Z}}{2}}.
\end{equation}
The additive constant term in the conditional probability %$\bar k g( d^2_{\mathbf Y | \mathbf x} + d^2_{\mathbf x}  )$ 
$(1+(d^2_{\mathbf Y }+ d^2_{\mathbf x} )/\nu)^{-\frac{\nu+p_{\mathbf Z}}{2}}$
can be handled so to keep the formula in the same form as Eq.\ref{g}, obtaining through simple algebra
\begin{equation}
f_{\mathbf Y | \mathbf x}(\mathbf Y) \! = \! \bar k  |\boldsymbol \Omega_{\mathbf Y\mathbf Y| \mathbf X}  |^{-1/2}  \left(1+ \frac{d^2_{\mathbf Y | \mathbf x}}{\nu_{\mathbf Y | \mathbf X}}\right)^{-\frac{\nu_{\mathbf Y | \mathbf X}+p_{\mathbf Y}}{2}}.
\end{equation}
where $\nu_{\mathbf Y | \mathbf X} = \nu + p_{\mathbf X}$ which depends on the dimension of $\mathbf X$ but not its values.
This is clearly still a multivariate Student-t with expectation values $ \boldsymbol\mu_{\mathbf Y| \mathbf x}$ and shape matrix 
\begin{equation}\label{St.Shape}
\boldsymbol \Omega^{St.t}_{\mathbf Y\mathbf Y| \mathbf x} 
=
\frac{\nu + d^2_{\mathbf x} }{\nu + p_{\mathbf X}} \boldsymbol \Omega_{\mathbf Y\mathbf Y| \mathbf X} .
\end{equation}
When defined, the conditional covariance is
\begin{equation}\label{e.Sproport}
\boldsymbol \Sigma^{St.t}_{\mathbf Y\mathbf Y| \mathbf x} 
= \frac{\nu + d^2_{\mathbf x}}{\nu + p_{\mathbf X}-2}\boldsymbol \Omega_{\mathbf Y\mathbf Y| \mathbf X} .
\end{equation}
therefore the covariance in the Student-t case is proportional to the conditional covariance of the multivariate normal distribution, however the extra coefficient in front is now dependent on the values of the conditioning variable $\mathbf x$. 

There are two considerations that need to be made here. 
First, one can note that the degrees of freedom becomes  $\nu_{\mathbf Y | \mathbf X} = \nu + p_{\mathbf X}$, which, from a systemic perspective, when $p_{\mathbf X}$ is large, means that the degrees of freedom becomes large and the conditional Student-t becomes closer to a multivariate normal. 
This effect reduces the probability of large deviations from the mean of the conditioned variable.
Second, the term $d^2_{\mathbf x}$ has expected value $\mathbb E[d^2_{\mathbf x}] = p_{\mathbf X}$ and it can vary in the range $[0,\infty)$; it can therefore have a very large impact on the conditional probability distribution and it can increase the probability of large deviations from the mean  of the conditioned variable.

Finally, let me note that, analogously to the Student-t bivariate case, and differently from normal modeling, the residuals $\mathbf Y -  \boldsymbol\mu_{\mathbf Y| \mathbf x}$ are not independent from $ \mathbf X$, however the scaled quantity $ \mathbf{\bar Y} = {\frac{\nu + p_{\mathbf X}}{\nu + d^2_{\mathbf x}}}(\mathbf Y -  \boldsymbol\mu_{\mathbf Y| \mathbf x})$ is instead independent from $ \mathbf X$.

\begin{figure}[h]
\begin{center}
\includegraphics[width=0.8\textwidth]{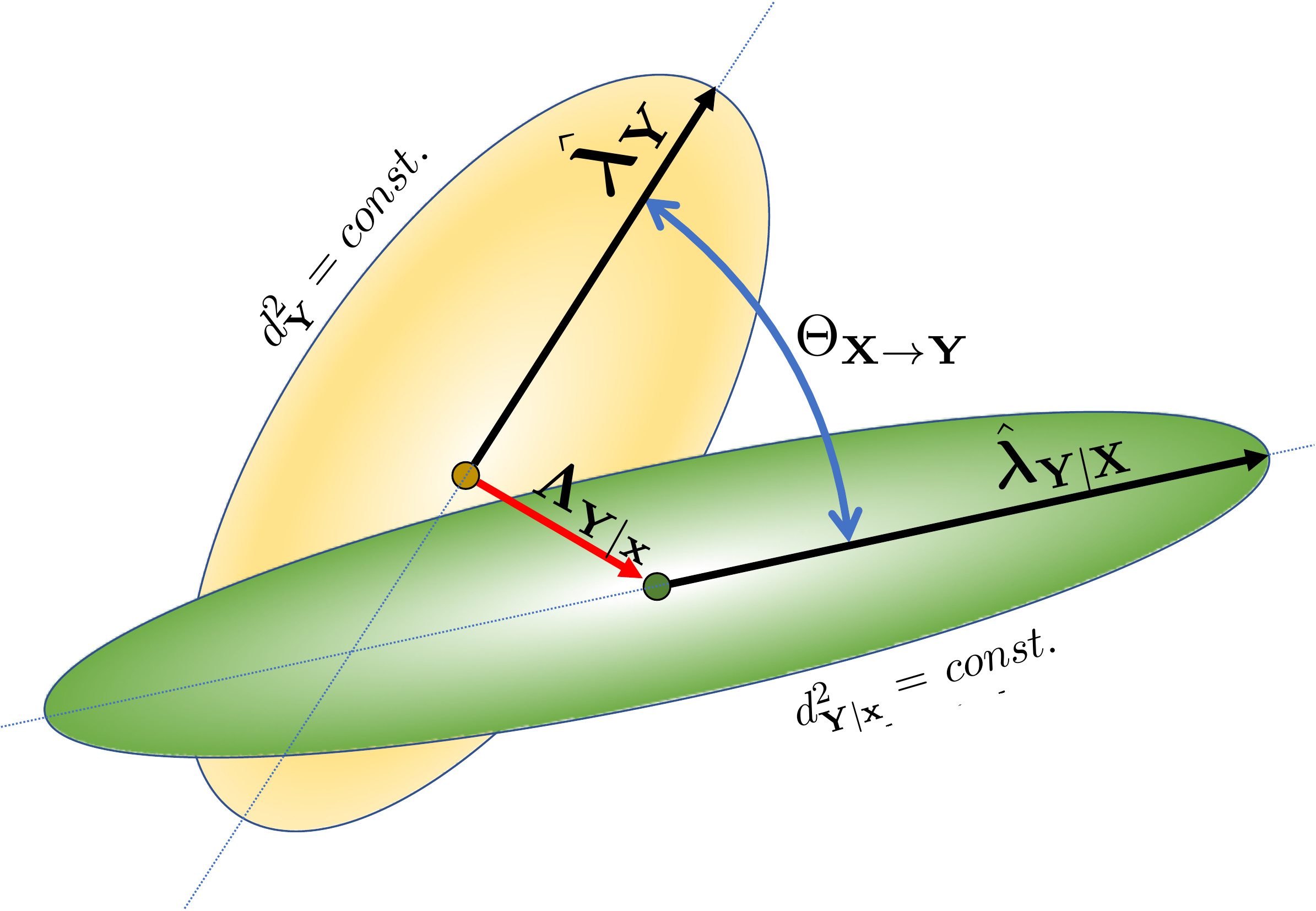}% This is a *.pdf file
\end{center}
\caption{
A pictorial representation of the effect of conditioning on the qui-probability surface of a multivariate elliptical distribution. 
Such surfaces are ellipsoids in a $p_{\mathbf Y}$-dimensional space. They are respectively described by the equations $d^2_{\mathbf Y } = const.$ and  $d^2_{\mathbf Y | \mathbf x} = const.$.
Conditioning to $ \mathbf X = \mathbf x$ shifts the barycentre of the ellipsoid $d^2_{\mathbf Y } = const.$  by the vector $\mathbf{\Lambda}_{\mathbf Y|\mathbf{x}}$, it rotates the most elongated axis by $\Theta_{\mathbf{X \to Y}} $ and it changes the length of such axis from ${\hat {\lambda}_{\mathbf Y}^{1/2}}$ to ${\hat {\lambda}_{\mathbf Y|\mathbf X}^{1/2}}$ }\label{f.Rotation}
\end{figure}

\section{Risk and stress measures for the multivariate elliptical family distributions} \label{s.2a}
 Let me now discuss some specific effects of conditioning on the properties of multivariate elliptical family distributions focusing on the few that, in my opinion, have the greatest relevance for the quantification of systemic risk and for stress testing. 
First, let me provide an intuitive vision of a multivariate distribution from the elliptical family from a geometrical perspective. 
The general expression $f_{\mathbf Z}(\mathbf{Z}) = k  |\mathbf{\Omega}|^{-1/2}  g(d^2_\mathbf{Z})$ in Eq.\ref{fZ} indicates that for the whole family the value of the probability density at any given point in space is determined by the  generalized Mahalanobis distance: two observations at the same Mahalanobis distance have the same probability density value.
From Eq.\ref{fd2}  one observes that the equation for the equi-probability region at constant Mahalanobis distance $d^2_\mathbf{Z}=const.$ corresponds to an ellipsoidal  surface in a $p_{\mathbf Y}$-dimensional space. 
Finally, from Eq.\ref{e.fYx} one observes that conditioning to $ \mathbf X = \mathbf x$ provokes a shift of the barycentre of the ellipsoid $d^2_{\mathbf Y } = const.$  by a vector $\mathbf{\Lambda}_{\mathbf Y|\mathbf{x}}$, it rotates the most elongated axis by an angle $\Theta_{\mathbf{X \to Y}} $ and it changes the length of such axis from $\hat {\lambda}^{1/2}_{\mathbf Y} $ to $\hat {\lambda}^{1/2}_{\mathbf Y|\mathbf X}$.
Figure \ref{f.Rotation} depicts such effect in a schematic way.

Let me now associate these major changes in the conditional probability density function with practical measures relevant for qualification of risk and stress testing.

\subsection{Shift of the centroid}
For the whole elliptical distribution family, the conditioning of a set of variables $\mathbf{Y}$ to another set of variables $\mathbf{X}$ produces as effect a shift in the position of the centroid of the probability distribution of the $\mathbf{Y}$ variables by the vector (see Eq.\ref{f.CondMU} and Fig.\ref{f.Rotation}) 
\begin{equation}
 \mathbf{\Lambda}_{\mathbf Y|\mathbf{x}}=  \boldsymbol{\mu}_{\mathbf Y|\mathbf x}  -  \boldsymbol{\mu}_{\mathbf Y} =  \mathbf{\Omega}_{\mathbf Y\mathbf X} \mathbf{\Omega}^{-1}_{\mathbf X\mathbf X} (\mathbf{x}- \boldsymbol{\mu}_{\mathbf X}).
\label{Shift}
\end{equation}
It must  be noted that, analogously to the uni-dimensional case, also in this case the shift in the locations coincide with the multilinear  regression of $\mathbf{Y}$ with respect to $\mathbf{X=x}$.  

As simple measure for stress test, I here propose to use the average extra losses on the subset $\mathbf{Y}$ provoked by extreme losses on the subset $\mathbf{X}$. 
This is in analogy with the reasoning beyond the CoVaR approach  \cite{adrian2011covar}, the idea is to stress the set of variables $\mathbf{X}$ to some extent and then measure the average losses caused in the $\mathbf{Y}$ set.
This is:
\begin{equation}
L_{\mathbf{X \to Y}}^q =  \frac{1}{  p_{\mathbf Y} } \mathbf{1_Y^\top} \mathbf{\Omega}_{\mathbf{YX}} \mathbf{\Omega}^{-1}_{\mathbf{XX}}
\mathbf{VaR_X}^q,
\label{e.L}
\end{equation} 
where $\mathbf{VaR_X}^q\in\mathbb R^{p_{\mathbf X}\times 1}$ is the vector of losses in $\mathbf X$ at value at risk $q$ and $\mathbf{1_Y}\in\mathbb R^{p_{\mathbf Y}\times 1}$ is a  column vectors of ones. 

Let me argue that, from a risk perspective this shift is the important factor because conditioning always reduces uncertainty.
This is discussed in  the next session. %\ref{A.UncertantyReduction}.

\subsection{Reduction of uncertainty}  \label{A.UncertantyReduction}
%Conditioning always reduces uncertainty.
Uncertainty associated with a set of variables $\mathbf{Z}$ can be quantified in terms of the Shannon entropy:
\begin{equation}
H(\mathbf{Z}) = - \mathbb E_\mathbf{Z}[\log f_\mathbf{Z}(\mathbf{Z})].
\end{equation}
With analogous definitions for $\mathbf{X}$ and $\mathbf{Y}$. 
The larger the entropy the greater is the uncertainty on the variables.
When the scale matrices are invertible, the explicit expressions for the elliptical family are \cite{arellano2013shannon}:
\begin{equation}
\begin{array}{ll}
H(\mathbf{Z}) = \frac{1}{2} \log(| \mathbf{\Omega}_{\mathbf Z\mathbf Z}|) + H_{Z_0}\\
H(\mathbf{X}) = \frac{1}{2} \log(| \mathbf{\Omega}_{\mathbf X\mathbf X}|) + H_{X_0} \\
H(\mathbf{Y}) = \frac{1}{2} \log(| \mathbf{\Omega}_{\mathbf Y\mathbf Y}|) + H_{Y_0} ,\\
\end{array}
\label{HHH}
\end{equation}
where  $H_{Z_0}$, $H_{X_0}$, $H_{Y_0}$ are the entropies associated with the standardized random vector with zero expected values and the identity as shape matrix. These quantities are therefore independent from the location vectors and the shape matrices.
The effect on uncertainty of $\mathbf{Y}$ caused by the conditioning to $\mathbf{X}$ can be quantified from the difference between the entropy of the unconditioned set of variables, $H(\mathbf{Y})$ and the conditioned one $H(\mathbf{Y}|\mathbf{X})$. 
It should be noted that $H(\mathbf{Y}|\mathbf{X})= H(\mathbf{X},\mathbf{Y}) - H(\mathbf{X})$ and therefore the difference is 
\begin{equation}
H(\mathbf{Y})-H(\mathbf{Y}|\mathbf{X})=H(\mathbf{X})+H(\mathbf{Y})-H(\mathbf{X},\mathbf{Y}) = I(\mathbf{X};\mathbf{Y})
\end{equation}
 which is the mutual information \cite{arellano2013shannon}. 
The mutual information $I(\mathbf{X};\mathbf{Y})$ is a non-negative quantity which is equal to zero when the two variables are independent. This tells us that conditioning always reduces uncertainty on the conditioned variable except in the case when the two sets are independent.
From the previous expressions in Eq.\ref{HHH} the mutual information is 
\begin{equation}
I(\mathbf{X};\mathbf{Y}) 
= \frac{1}{2} \log(\frac{| \mathbf{\Omega}_{\mathbf X\mathbf X}|| \mathbf{\Omega}_{\mathbf Y\mathbf Y}|}{| \mathbf{\Omega}_{\mathbf Z\mathbf Z}|}).
\label{IXY}
\end{equation}
In the present context, the mutual information is interpreted as a measure of reduction of uncertainty on the variable $\mathbf Y$ deriving from constraining $\mathbf X$ at some stressing value. Such a reduction is not dependent on the value of variable $\mathbf X$. 

This measure must be interpreted in combination with the previous. A larger mutual information  means a greater reduction in uncertainty deriving from constraining but not necessarily a reduction in risk which is mainly carried by the shift of the centroid. 
Furthermore, as we shall see shortly in the next subsection, even if uncertainty is always reduced by constraining, variability might increase instead when non-normal models are considered.

{\color{black}
Let me note that the conditional entropy $H(\mathbf{Y}|\mathbf{X})$  is the entropy  associated with the shifted and scaled variable  $\mathbf{\bar Y} = {\frac{1}{\nu + d^2_{\mathbf x}}}(\mathbf Y -  \boldsymbol\mu_{\mathbf Y| \mathbf x})$. 
It is not dependent on the values of $\mathbf{X}=\mathbf{x}$ and it is not given by $ - \mathbb E_\mathbf{Y|X}[\log f_\mathbf{Y|x}(\mathbf{Y})] $ 
which is instead a function of $\mathbf{x}$.
}

In terms of risk and stress testing, we have two opposite factors. The more two parts of the system are dependent, the larger will be the effect of one part on the shift of the centroid of the other part. This imply an increase in risk. 
However, the same large dependency between the variables cause a large reduction in uncertainty  under conditioning. 
Therefore, one has two competing mechanisms:   risk increases with dependency due to the shift of the centroid but it decreases with dependency due to the reduction in uncertainty.

\subsection{Rotation of the  principal axis}\label{s.Rotation}
The eigenvector $\hat {\mathbf u}_{\mathbf Y | \mathbf X}$ associated with the largest eigenvalue of  the conditional covariance is not the same as the eigenvector $\hat {\mathbf u}_{\mathbf Y }$  associate with the largest eigenvalue of the to the unconditional covariance.
They are respectively the eigenvectors of  $\boldsymbol \Omega_{\mathbf Y\mathbf Y| \mathbf X}$ and   $\boldsymbol \Omega_{\mathbf Y\mathbf Y}$
They are both unitary in module and therefore the change is a rotation by the angle:
\begin{equation}
\Theta_{\mathbf{X \to Y}} = \arccos( \hat {\mathbf u}_{\mathbf Y}^\top \hat {\mathbf u}_{\mathbf Y|\mathbf X}).
\label{e.T}
\end{equation} 
This rotation of the axis of largest elongation of the equi-probility elliptical surface provides information on relative increase or decrease of risk associated with each variable.
Under stress, some variables might take larger weights than in normal unstressed situations highlighting systemic fragility.  
The larger the rotation the larger is the disruptive effect of conditioning on the variables.
I therefore propose that this angle is a simple measure of systemic effect of the stress. 
Such effect is not directly a measure of risk because this will depend on the assets in the investors' portfolios. However it implies a change in the market conditions and must be taken into account for risk management purposes.

These eigenvectors are the same for the whole family of elliptical distributions. 
Indeed, the whole family share the same shape matrices that might differ only by a scalar factor.  
Therefore, this is also a general measure of risk  independent from  modeling details. 
Note that they do not depend on the values of the stressing variables.

There are many other possible measures of the effect of conditioning on the statistical properties of the system.  
Some of them are mentioned in \ref{A.VariabilityFactors}. 
Specifically, one can look  at the change in the size of the axis (\ref{A.AxSiz}); at the change in the variance of an arbitrary portfolio (\ref{A.VarPort}) and; at the change in total variance (\ref{A.TotVar}).
All these measures are not independent and indeed most of them align with the mutual information measure.
Differently from the measures just discussed in this section, the other measures reported in the appendix are not universal for the entire elliptical family. I will therefore discuss them for the two cases of the multivariate normal and Student-t.

\begin{figure}[h]
\begin{center}
\includegraphics[width=0.8\textwidth]{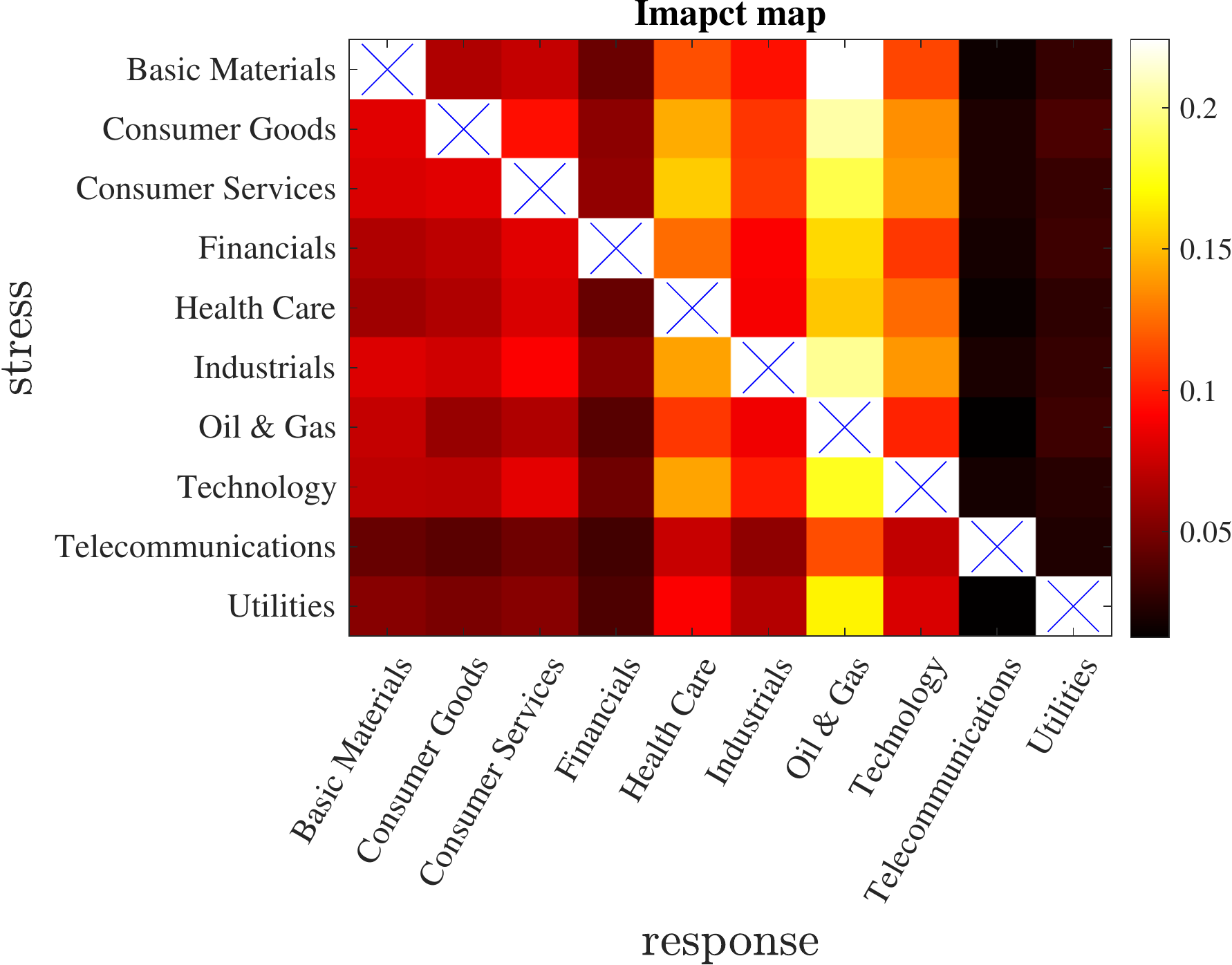}% This is a *.pdf file
\end{center}
\caption{
$L_{\mathbf{X \to Y}}^{0.95}$: average losses in a sector as consequence of conditioning another sector. 
The color map corresponds to the average amount of losses (negative log-returns) on a the sector reported on the columns (conditioned variables) caused by stressing at 95\% VaR the sector reported on the rows (conditioning variables). 
}\label{ImpactMap}
\end{figure}

\begin{figure}[h]
\begin{center}
\includegraphics[width=0.8\textwidth]{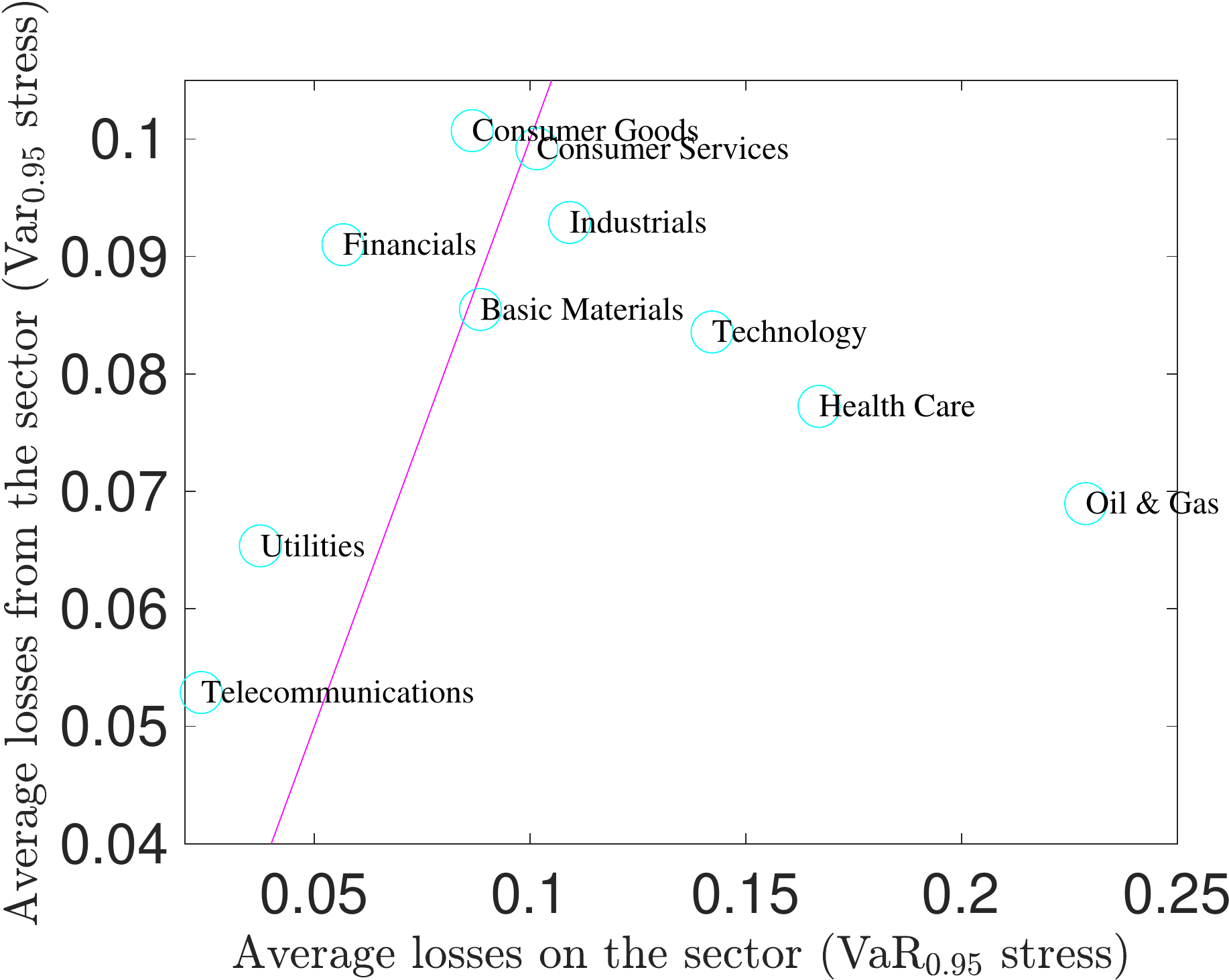}% This is a *.pdf file
\end{center}
\caption{Average losses caused on the entire system by stressing a given sector to 95\% of its VaR vs. the losses on the sector when the rest of the system is stressed to 95\% VaR.
The line is the diagonal separation  and is reported for visual reference. }\label{ImpactPlot}
\end{figure}

\section{Stress testing experiment with US equity market}
\label{s.4}
To provide an example of stress testing I collected data for 623 equities continuously traded on the US market between 01/02/1999 and 07/04/2020 for a total of 5329 trading days.
I computed the log-returns of daily prices and performed the analysis as described in the previous sessions.
In particular, I computed the centroid shift $L_{\mathbf{X \to Y}}^{0.95}$, the reduction in uncertainty quantified by $I(\mathbf{X};\mathbf{Y})$ and the change in risk distribution associated with the rotation $\Theta_{\mathbf{X \to Y}}$.
The focus in this paper is on the validation of these three risk measures by uncovering some expected spillover mechanisms. Once validated, future works will be instead focused in uncovering unexpected mechanisms.
Another reason for this experiment with real data is to compare these measures among themselves and assess if they are conveying equivalent or complementary information.

Results for $L_{\mathbf{X \to Y}}^{0.95}$ are reported in Fig.\ref{ImpactMap} where the columns report in color map the  average amount of losses in the relative sector caused by stressing the sectors reported on rows. % columns -> rows !!!
For instance, I measure 22\% average losses on oil and gas caused by  stressing the basic materials industry sector. 
Whereas, in the other direction,  stressing on oil and gas  causes only 7\% losses on basic materials.
Note that this is not the CoVaR, it is the mean that is moved by this quantity.  
A shift of 22\% of the mean indicates that without any endogenous stress, the exogenous stress propagating from the basic materials  sector causes already considerable losses in the  oil and gas sector from close to zero to 0.22.  
It is evident from this example that the effects of a sector on another are not symmetrical.
It is also noticeable that telecommunications and utilities are little influenced by stress on other sectors wheres  oil and gas, healthcare and technology are much more influenced. 

I computed also the effect of each industry sector on the rest of the market and vice-versa. The results are reported in Fig.\ref{ImpactPlot}. 
One can note that, oil and gas, healthcare and technology sectors are indeed the most impacted by stressing the rest of the market while consumer (goods and services), financials and industrials are the sectors that impact most the whole market. 
Note that the results in Fig. \ref{ImpactPlot} are similar to but not coincident with what it would be obtained by plotting on the vertical axis the mean of the rows and on the horizontal axis the mean of the columns of the matrix in Fig.\ref{ImpactMap}. 
Indeed, the quantity reported in Fig. \ref{ImpactPlot} is a multivariate estimate of the effects of a group of sectors onto another and it takes into account all the internal effects of attenuation and sometimes amplification of the propagating stress. 

Let me note that,  in some analogy with the CoVaR approach, I stressed the variables set $\mathbf{X}$ at their 95\% VaR. 
However, this is relatively arbitrary way to set stressed values.
I verified that other levels of VaR give similar results, for instance stressing the more extreme 99\% VaR produces very similar outcomes though with roughly doubled values of average losses. 
Another possibility is to stress with a uniform stress of, for instance, all equal to 1 (the number itself is irrelevant because the results scale linearly with it).
Results in this case are consistent with the previous but quantitatively quite different. It is beyond the purpose of the present work to investigate which kind of stressing is the most adequate and, most likely, this depends on the purpose of the stress testing exercise.

\begin{figure}[h]
\begin{center}
\includegraphics[width=0.8\textwidth]{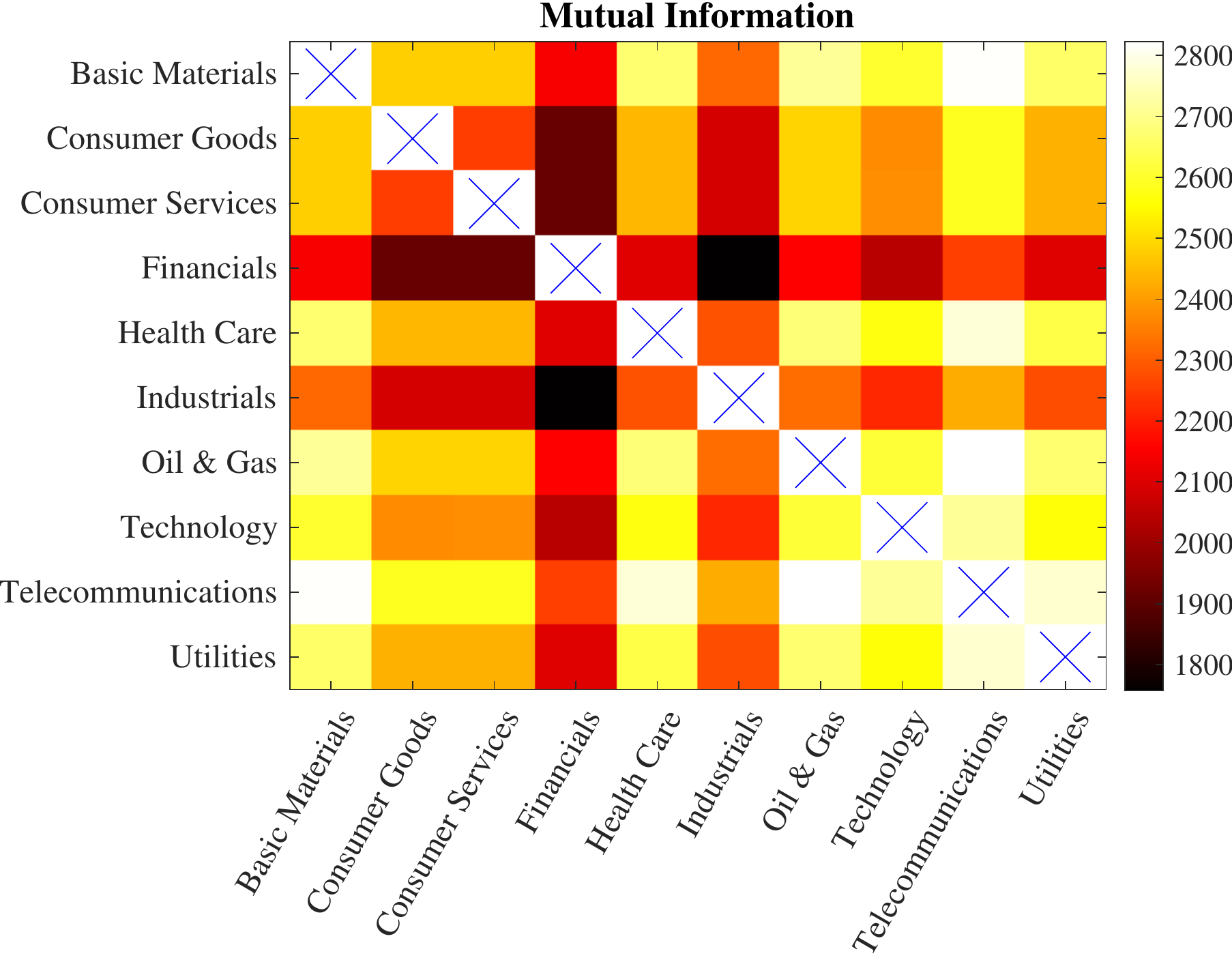}% This is a *.pdf file
\end{center}
\caption{
Intensity map for the mutual information, $I(\mathbf{X};\mathbf{Y})$, between industry sectors.
}\label{MI}
\end{figure}

As discussed in the previous section, while losses are induced by the  shift of the centroid, uncertainty is instead always reduced by conditioning and endogenous losses under conditioning are lower than expected in the unconstrained case.
The effect of uncertainty reduction and consequently of conditioning is illustrated in Fig.\ref{MI} where the values of the mutual informations, $I(\mathbf{X};\mathbf{Y})$, are reported for the various industry sectors.
Note that, in this case, the matrix is symmetric and the results do not depend on the values of $\mathbf{X}$ and its level of stress.
Interestingly, the map is quite different from the one for the stress propagation in Fig.\ref{ImpactMap}. 
Note, for instance, that telecommunications and utilities have large levels of mutual information while had small induced losses. 
This is confirmed by looking at correlations between the values of the two measures $ \mathbf{\Lambda}_{\mathbf Y|\mathbf{x}}$ and $I(\mathbf{X};\mathbf{Y})$ across all the cross-sectors which does not reveal overall significant overlapping patterns.

\begin{figure}[h]
\begin{center}
\includegraphics[width=0.8\textwidth]{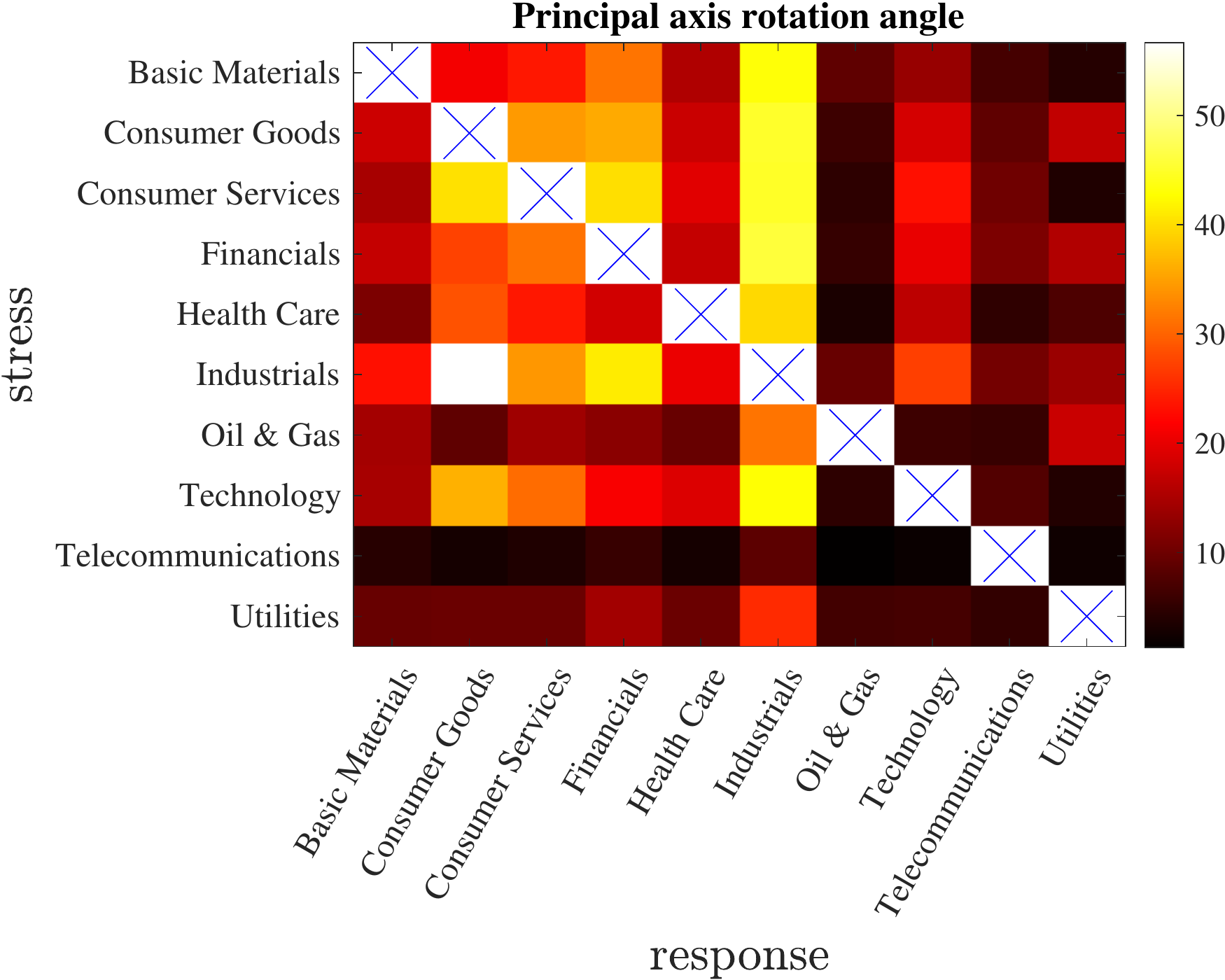}% This is a *.pdf file
\end{center}
\caption{Angles of rotation, $\Theta_{\mathbf{X \to Y}}$, of the principal axes as consequence of conditioning (degrees). The rows are the conditioning variables and the columns the conditioned ones.}\label{AnglePlot}
\end{figure}

The angle of rotation of the principal axis, $\Theta_{\mathbf{X \to Y}}$, is reported in Fig.\ref{AnglePlot}. 
I note that the effect of conditioning a sector over the other sectors can be quite severe on the change in direction of the principal axis. 
I observe, for instance, that the industrial sector is highly affected by stressing and technology and consumer services tend to strongly affect a number of other sectors. 
Telecommunications is the least influencing while, contrary to the results for $L_{\mathbf{X \to Y}}^{0.95}$, under this measure, oil and gas is the least influenced. 
These rotations are not directly a measure of losses, they indicate systemic effects of a sector onto another sector measuring the change in relevance of the contribution of some variables towards the principal axis of variability.
By looking at the relation with the previous measures I observe that this measure has small overlaps with $L_{\mathbf{X \to Y}}^{0.95}$ with no significant correlation, while instead it reflects well the mutual information with a strong negative correlation around 70\%.

\section{Conclusions}
\label{s.5}
The effects of stress testing  a multivariate system of interacting variables are described by the conditional probability distribution where all statistical properties of the stressed variables can be expressed as function of the values of the stressing variables. 
In this work I have shown that for the vast class of multivariate models described by elliptical distributions, the conditional probability density function is strictly linked to the original unconditional probability.
I pointed at three main effects of conditioning affecting the elliptical equi-probability surface: 1.  shift in the centroid; 2. reduction in uncertainty, and; 3.rotation of the axis.  
I have introduced three measures that quantitatively captures these three effects.
I have argued that the shift in the centroid is the primary cause of stress and loss propagation showing that in general the uncertainty on the conditioned variables is reduced by conditioning. 
Specifically, for multivariate normal modeling the shift is the centroid is the only possible source of extra losses induced by stressing part of the system. 
Instead, for Student-t there could potentially be other amplification effects. 
The centroid shift  coincides with the multivariate linear regression factor and therefore strongly dependent variables propagate stress more  than weakly dependent ones. However, the opposite is instead true for what concerns uncertainty and endogenous losses: the larger is the dependency and the smaller is the residual uncertainty on the constrained variables. 

Application to an example of equity returns shows meaningful outcomes. 
The three measures reveal consistent but different mutual effects between industry sectors. 

The proposed risk measures are quite intuitive and easy to implement. The elliptical distribution family includes the multivariate Student-t distributions that are quite realistic models for risk in financial systems.    
The practical limitation of this approach is the estimate of the parameters of the multivariate distribution and in particular the shape matrix $ \mathbf{\Omega}$  which comprises $p(p-1)/2$ parameters and therefore requires a large number of consistent observations to be properly estimated.
However, this limitation can be dealt with by using recently developed techniques that allow $L_0$ topological regularization using network filtering techniques \cite{LoGo16} which overcome the curse of dimensionality problem.
Stress testing using specific market states can be performed by using the clustering method presented in \cite{procacci2019forecasting} which also  overcome  non-stationarity issues. 
All these measures introduced in this paper and a test dataset are accessible for public use at: \emph{https://github.com/financialcomputing/Systemic-Risk-Measures}.

Despite the elliptical family being rather vast, one of its shortcomings is that it is symmetric with positive and negative fluctuations having the same statistics. It is instead well established that the statistics of financial returns is not symmetric. 
An extension to non-symmetric multivariate distribution is the multivariate generalized hyperbolic (MGH) family that has been indeed indicated as particularly adequate for the modeling of financial returns \cite{Prause99}. 
The approach presented in this paper is focused around the effect of conditioning on the vector of expected values and on the $d^2_{\mathbf X}=const.$ ellipsoid. 
These aspects stay unchanged in the MGH modeling, however other factors and parameters come into paly as well. 
Extending the present study to the  MGH family is a meaningful  continuation of this work.

\section*{Acknowledgments}\noindent
The author acknowledges discussions with many members of the Financial Computing and Analytics group at UCL. In particular a special thank to Fabio Caccioli, Guido Massara, Carolyn Phelan and Pier Francesco Procacci.  
Also, thanks for support from ESRC (ES/K002309/1),  EPSRC (EP/P031730/1) and EC (H2020-ICT-2018-2 825215).

\appendix

\section{Other measures for systemic risk and stress testing}
\label{A.VariabilityFactors}
There are several possible measures that can be used to capture systemic risk and the effects of stress in multivariate systems. 

\subsection{Change in principal axis  size } \label{A.AxSiz}
Together with the rotation discussed in the subsection \ref{s.Rotation}, the principal axis also change in size.
The length of the largest axis is the square root of the largest eigenvalues of the unconditional and conditional covariance matrices, which I denote  $\hat {\lambda}_{\mathbf Y}$ , $\hat {\lambda}_{\mathbf Y|\mathbf X}$ respectivelly.
A simple measure is the relative change in value for the constrained variables with respect the unconstrained one:
\begin{equation}
\Delta_{\mathbf{X \to Y}} = \frac{ \hat {\lambda}_{\mathbf Y}  - \hat {\lambda}_{\mathbf Y|\mathbf X} }{ \hat {\lambda}_{\mathbf Y} }.
\label{e.D}
\end{equation} 
The largest eigenvalue is associated with the most elongated principal axis of the shape matrix  and it is proportional to the variance of the  linear combination $\tilde Y = \hat {\mathbf u}^\top \mathbf Y$: the portfolio of the `riskiest' component.

These are the eigenvalues of the covariance matrix, they are directly proportional  to the eigenvalues of the shape matrix.
The coefficients of proportionality depend on the kind of elliptical distribution assumed in the model.
For the multivariate normal distribution the covariance and shape matrices coincide whereas for the Student-t the constant of proportionality is $\frac{\nu + d^2_{\mathbf x}}{\nu + p_{\mathbf X}-2}$ (see Eq.\ref{e.Sproport}).

We shall see later, with real data experiments, that the size of the principal axis shrinks under conditioning. 
This is intuitive because, as pointed out in section \ref{A.UncertantyReduction}, uncertainty decreases under conditioning.

{
\subsection{Change in the variance of an arbitrary portfolio} \label{A.VarPort}

For risk and stress testing purposes the relevant measure is the likelihood of losses for a given set of assets in a portfolio.
Of course, this quantity depends on the portfolio weights and indeed the main task in portfolio management is to optimize the choice of such  weights.
The average losses are simply related to the shift in the centroid. 
For a given portfolio $\tilde Y = \mathbf w^\top \mathbf Y$,  the average losses caused by a conditioning to $\mathbf{X=x}$ are: $\mathbf w^\top \mathbf{\Lambda}_{\mathbf Y|\mathbf{x}}$. 
This is universal for the whole elliptical family distribution. 
Conversely, the changes of the variance of the portfolio  is not general and it depends on the kind of elliptical distribution one considers. 
Indeed, generally speaking, for the elliptical family distribution the likelihood of large variations is directly related to the covariance and the covariance is proportional to the shape matrix with a proportionality constant that depends on the kind of distribution.

The {  variance} of any conditioned linear combination of variables $\tilde Y = \mathbf w^\top \mathbf Y$ is:
\begin{equation}
\sigma^2_{ \tilde Y | \mathbf X} =\beta_{\mathbf x} 
\left(
%\frac{\nu-2}{\nu}
\sigma^2_{ \tilde Y} -  
\mathbf w^\top \boldsymbol \Omega_{ \mathbf Y \mathbf X}  \boldsymbol \Omega_{\mathbf X \mathbf X}^{-1} \boldsymbol \Omega_{\mathbf X  \mathbf Y } \mathbf w
\right) .\label{s2Y|X}
\end{equation}
where $ {\mathbf w}^\top \boldsymbol \Omega_{ \mathbf Y \mathbf X}  \boldsymbol \Omega_{\mathbf X \mathbf X}^{-1} \boldsymbol \Omega_{\mathbf X  \mathbf Y }  {\mathbf w}\ge 0$ for any ${\mathbf w}$ and %$\beta_{\mathbf x}=1$ in the normal case but in the Student-t case
\begin{equation}\label{Delta0}
\beta_{\mathbf x} = \begin{cases} \displaystyle {1} & \text{Normal case;} \\ {\displaystyle \frac{\nu + d^2_{\mathbf x}}{\nu + p_{\mathbf X} -2}} & \text{Student-t case}. \end{cases}
\end{equation}
Therefore, for the multivariate normal case the variance of any arbitrary portfolio is always reduced by conditioning.
Conversely, this is not the case for the Student-t where the multiplicative factor $\beta_{\mathbf x}$ can increase variability. 
This factor multiplies the shape matrix (see Eq.\ref{e.Sproport}) and therefore it affects all risk measures associated with the covariance.
Let me note that, when the  variables $\mathbf X$ are set at their expected values $\mathbf x = \boldsymbol \mu_X$, then $d^2_{\mathbf x}=0$. 
This is a `zero stress' condition and the factor is smaller than one. 
However, $\mathbb E(d^2_{\mathbf X}) =  p_{\mathbf X}$ and one expects $d^2_{\mathbf x} > p_{\mathbf X}$ when the stress level is above the mean.
In general, the distance $d_{\mathbf x}$ measures how distant a datapoint is from the equi-probability elliptical surface which contains the body of the elliptical distribution.  
Any stressing variable is likely to have values outside such an envelop.
We can see from Eq.\ref{e.fYx} that $d^2_{\mathbf x} $ plays a very important role as an additive factor to the generalized Mahalanobis distance in the expression for the conditional probability density for the whole elliptical family. We have seen that the effect of this term is different for each specific distribution and, in the Student-t case, it lead to the multiplicative factor $\beta_{\mathbf x}=\frac{\nu + d^2_{\mathbf x}}{\nu + p_{\mathbf X} -2}$. 
The multivariate nature of the problem makes very hard to identify a simple scalar measure of stress impact in this context.
I argue that, in general the relative value of  $d^2_{\mathbf x}$ with respect to $p_{\mathbf X}$ is a good measure of extra risk factor in non-normal modeling. 
Therefore, I propose the following measure:
\begin{equation}
B_{\mathbf{x}} = \frac{d^2_{\mathbf x^*_q} }{ p_{\mathbf X} },
\label{e.B}
\end{equation} 
where, in some analogy with the CoVar approach I propose to assign to $d^2_{\mathbf x^*_q}$  the value of the top $q$-quantile of ${\mathbf X}$.
Values of $B_{\mathbf{x}} $ significantly larger than one are expected to significantly increase the effect of conditioning on risk in the Student-t case and in general when the system does not follow multivariate normal statistics.
This factor depends only on the stressing variables $\mathbf X$ and quantify the relative impact of subset of variables on the other.
I call this quantity the {\bf Mahalanobis impact factor}.
}

\subsection{Total variance change} \label{A.TotVar}
{\color{black}
The total variance is defined as the determinant of the covariance matrix and it is a global measure of variability; it quantifies the overall occupied volume in the probability phase space. 
The change in the total variance as consequence of conditioning is a measure of change in global variability.
Such a change depends  on the modeling. Specifically:
\begin{equation}\label{SY}
|\boldsymbol \Sigma_{\mathbf Y\mathbf Y| \mathbf x} | =  
\begin{cases} 
\displaystyle {|\boldsymbol \Omega _{\mathbf{YY|X}} | } & \text{Normal case;} \\ 
\displaystyle{\left( \frac{\nu  + d^2_{\mathbf x}}{ \nu + p_{\mathbf X} -2} \right)^{p_{\mathbf Y}}|\boldsymbol \Omega _{\mathbf{YY|X}} | }   & \text{Student-t case}. \end{cases}
\end{equation}
Constraining always reduces the determinant and therefore:
\begin{equation}
|\boldsymbol \Omega_{\mathbf{YY|X}} | \le |\boldsymbol \Omega_{\mathbf{YY}}| .
\end{equation}
However, as a consequence of the factor in front, for Student-t multivariate modeling the total variance of the conditioned variables, $|\boldsymbol \Sigma_{\mathbf Y\mathbf Y| \mathbf x} |$, could in principle increase with respect to the total variance in the unconditioned case, $|\boldsymbol \Sigma_{\mathbf Y\mathbf Y} |$.
Indeed, when $d^2_{\mathbf x} > p_{\mathbf X}-2$, the total variance can increase, it will instead decrease when $d^2_{\mathbf x} < p_{\mathbf X}-2$.
The presence of $p_{\mathbf Y}$ at the exponent in the Student-t case makes the effect of the multiplicative factor much stronger for large $p_{\mathbf Y}$. 

%I have shown that for all these risk measures in the non-linear modeling case with the Student-t distribution the square Mahlanobis distance $d^2_{\mathbf x}$ plays an important role.  
%Specifically, $d^2_{\mathbf x}> p_{\mathbf X}$ can cause a systemic risk increase amplifying the effect of the stressed variable.
%I therefore propose to 
%I call this quantity the {\bf Mahalanobis impact factor}, indeed I have shown that this factor has an important role in non-linear modeling, for instance in the Student-t case.

\section{Experiment results on the other stress measures}\label{A.b}

\subsection{Change in the eigenvectors' length}
The change in the length of the principal axis, $\Delta_{\mathbf{X \to Y}}$ (Eq.\ref{e.D}), is reported in Fig.\ref{LenghtStretchPlot}. 
This measure is significantly correlated with the mutual information with about negative 60\% correlation across sectors. 
The negative correlation indicates that larger dependency has as a consequence smaller variability.
\begin{figure}[h]
\begin{center}
\includegraphics[width=0.8\textwidth]{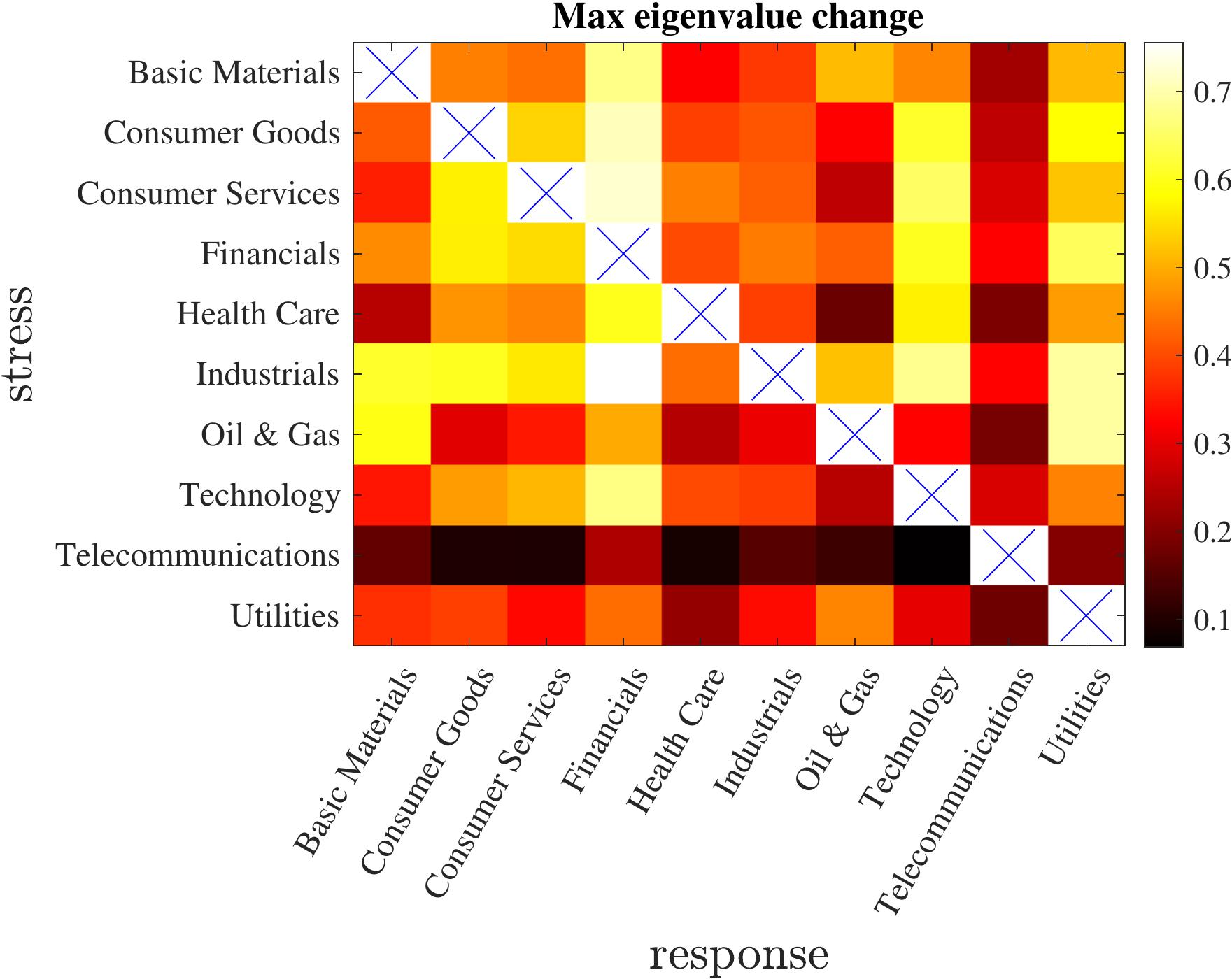}% This is a *.pdf file
\end{center}
\caption{Change in the length of the principal axis, $\Delta_{\mathbf{X \to Y}}$}\label{LenghtStretchPlot}
\end{figure}

\subsection{The Mahalanobis impact factor}
The Mahalanobis impact factor $B_{\mathbf{x}} = d_{\mathbf x^*_q}/p_{\mathbf X}$ (Eq.\ref{e.B}) is reported in in Fig.\ref{dx}.  
%it is indeed the factor that appears in the coefficient in front of the covariance 
It results  that $d_{\mathbf x^*_q} < p_{\mathbf X}$ for all sectors with the largest effect  from the  Telecommunication sector. 
This means that for this example also for the Student-t case stressing a sector always reduces the variability on the other sectors.
This is quite remarkable, given that ${\mathbf X}$ are stressed at their 95\% VaR, but indeed it turns out that values of $d_{\mathbf x^*_q} > p_{\mathbf X}$ are recorded only for stresses above the 99\% quantile. 
Note that the values reported are computed using the covariance and therefore are the coefficients for the multivariate normal case. 
For the multivariate Student-t case they will be further reduced by a factor $(\nu-2)/\nu$.

\begin{figure}[h]
\begin{center}
\includegraphics[width=0.8\textwidth]{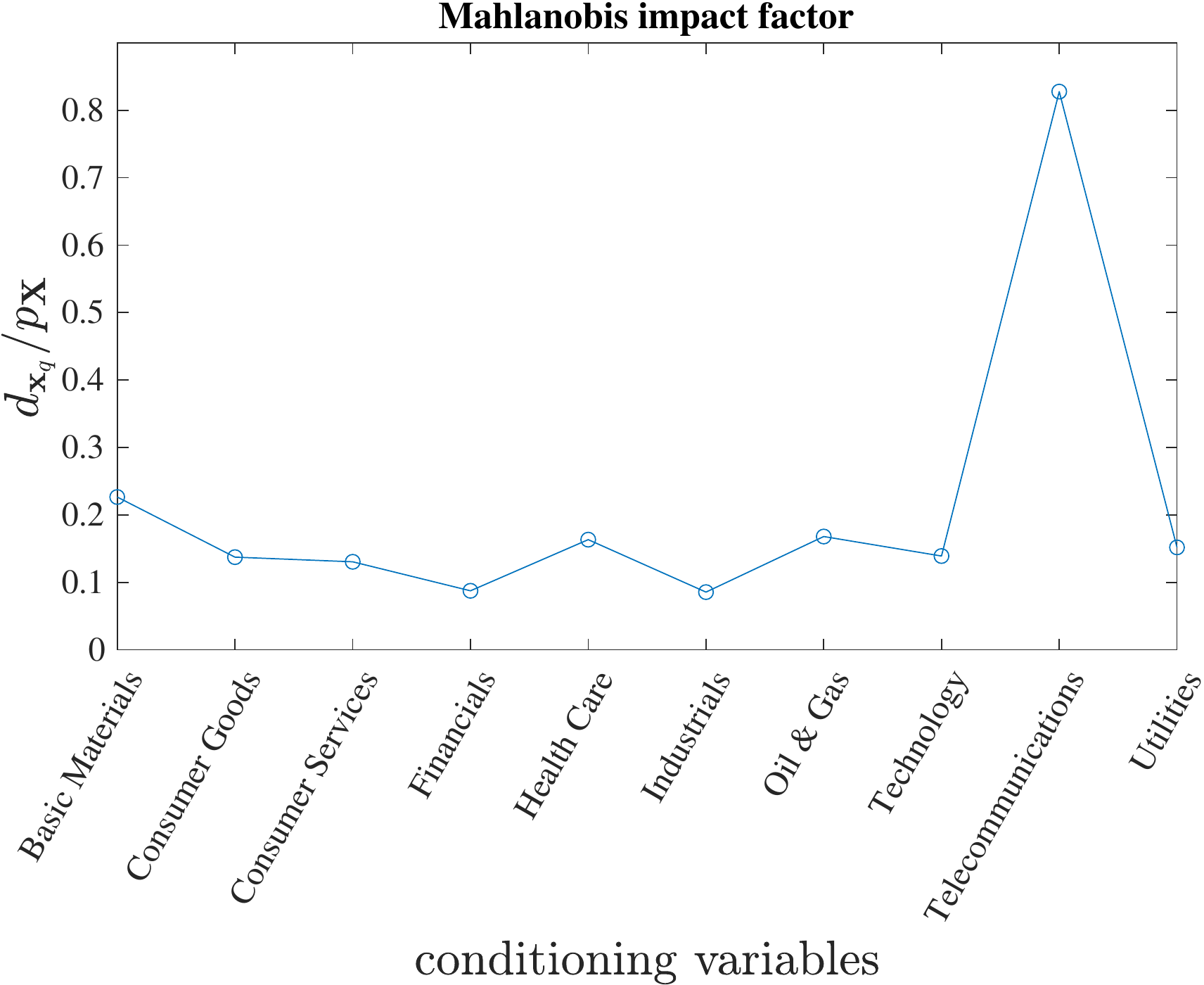}% This is a *.pdf file
\end{center}
\caption{Value of $B_{\mathbf{x}} = d_{\mathbf x^*_q}/p_{\mathbf X}$ (for the multivariate Normal case) for $q=95$\%.}\label{dx}
\end{figure}

%\section*{References}
\bibliographystyle{unsrt}
%\bibliography{StressTestMultivar.bib}
%\newpage
%\begin{appendix}
%{\Large{\textbf{Appendix}}}
%\end{appendix}

\end{document}